\documentstyle[11pt, aaspp4] {article}
\def\etal{\it et al.~\rm}
\begin{document}
\title{Weak Lensing by Two $z \sim  0.8$ Clusters of Galaxies}
\author{D. Clowe\altaffilmark{1}, G.A. Luppino, N. Kaiser, J.P. Henry
and I.M. Gioia\altaffilmark{2}}
\affil{Institute for Astronomy, University of Hawaii, 2680 Woodlawn Drive, 
Honolulu, HI 96822}
\altaffiltext{1}{Visiting Astronomer at the W. M. Keck Observatory, jointly 
operated by the California Institute of Technology and the University of 
California}
\altaffiltext{2}{Also Istituto di Radioastronomia del CNR, Via Gobetti 101, 
40129 Bologna, Italy}

\begin{abstract}
We report the detection of weak gravitational lensing of distant background
galaxies by two rich, X-ray luminous clusters of galaxies, MS 1137$+$66 at
$z = 0.783$ and RXJ 1716$+$67 at $z = 0.813$.  We detect a shear signal in
each cluster over the radial range $100 \leq  r \leq $ 700  $h^{-1}$ kpc.
Assuming the background galaxies lie in a sheet at $z=2$, we measure a 
mass of $2.45 \times  10^{14} h^{-1} M_\odot$ and $M/L_V = 270$ at 
$500 h^{-1}$ kpc radius for MS 1137$+$66, and a mass of $2.6 \times  10^{14}
h^{-1} M_\odot$ and $M/L_V = 190$ for RXJ 1716$+$67 at the same radius.  
Both the optical lightmaps and weak lensing massmaps of
RXJ 1716$+$67 show two spatially distinct sub-clusters,
as well as a long filamentary structure extending out of the cluster to the
northeast. In contrast,
MS 1137$+$66 is an ultra-compact massive cluster in
both massmaps and optical lightmaps, and contains a strongly-lensed arc
system in the cluster core.  These data add to the growing number
of massive clusters at $z \gtrsim  0.8$.
\end{abstract}


\section{Introduction}
The existence of high-mass, X-ray luminous clusters of galaxies at $z > 0.5$ 
provides a strong constraint on cosmological models (Eke \etal 
\markcite{r4} 1996).
The presence of any $\sim 10^{15} M_\odot$
clusters at $z\sim 0.8$ invalidates most $\Omega_0 = 1, \Lambda = 0$ CDM
models (Bahcall {\it et al.} \markcite{r1} 1997, Henry \markcite{r17} 1997).  
Further, the surface densities, amounts 
of sub-clustering, and mass to light ratios for these high $z$ clusters can
place even stronger constraints on cosmological models and the nature of
dark matter (Wilson \etal \markcite{r16} 1997, Crone \etal
\markcite{r3} 1996).

Currently, X-ray observations provide the most effective way of finding massive
clusters at high redshifts.  The Einstein Extended Medium Sensitivity Survey
(EMSS) and the ROSAT North Ecliptic Pole (NEP) survey have, so far, 
found three $z\sim 0.8$ clusters of galaxies (Gioia and Luppino 
\markcite{r18} 1994, Henry \etal \markcite{r8} 1997).  
One of these clusters, MS 1054$-$03, has had its mass confirmed by weak
gravitational lensing measurements (Luppino and Kaiser \markcite{r11} 1997,
hereafter LK).  
In this paper we report on the detection of weak gravitational
lensing signals in two more $z\sim0.8$ clusters of galaxies, MS 1137$+$66 and 
RXJ 1716$+$67.

MS1137$+$66 is one of the high redshift clusters from the EMSS 
(Luppino and Gioia \markcite{r7} 1995).  
It has a redshift of $z = 0.782$ and an x-ray luminosity of 
$L_x^{0.3-3.5 keV} = 1.9\pm0.4 \times 10^{44} h^{-2}$ erg s$^{-1}$.
RXJ 1716$+$67 is a recently discovered high-redshift cluster from
the ROSAT NEP survey.  It 
has a redshift of $z = 0.813$ and an x-ray luminosity of
$L_x^{0.3-3.5 keV} = 1.5\pm0.4 \times 10^{44} h^{-2}$ erg s$^{-1}$
(converted from the 0.5-2.0 keV ROSAT data assuming a $T_X =$ 6 keV).
The X-ray luminosities of these clusters are comparable to MS 1054$-$03
and suggest they may also be massive systems.  

Our observations, data reduction, and the weak lensing analysis are outlined 
in \S 2.  In \S 3 we discuss the properties of the cluster and the weak
lensing results.  Section 4 contains our conclusions.  We assume an
$\Omega _0=1, \Lambda =0$ cosmology throughout this paper, and all errors
are $1\sigma $.

\section{Observations, Data Reduction, and Weak Lensing Analysis}
Very deep
optical $R$-band images of RXJ 1716$+$67 and MS 1137$+$66 were obtained
with the Keck II 10m telescope on the nights of August 17-18 1996 and
January 10-11, 1997.  The images were obtained using LRIS (Oke \etal
\markcite{r13} 1995) in imaging
mode, resulting in a scale of 0\farcs215 pixel$^{-1}$ and a field of view
of 6\arcmin $\times$ 8\arcmin.  Total exposure times were 7500 s and 8700 s
respectively.

Optical $I$-band images were obtained with the UH 2.2 m telescope on the 
nights of July 20-22, 1996 for RXJ 1716$+$67 and April 6-7, 1995 and April 29 
- May 1, 1997 for MS 1137$+$66.  A thinned Tek 2048$^2$ CCD was mounted at 
the f/10 RC focus resulting in a scale of 0\farcs22 pixel$^{-1}$ and a field
of view of 7\farcm5 $\times$ 7\farcm5.  Total exposure times were
26100 s and 21000 s respectively.

The individual images in each filter were first debiased and then flattened
using a median of all the images taken in that filter during each set of
observations.  The $I$-band 88'' images were shifted into registration
and summed with cosmic-ray rejection.
The $R$-band Keck images were re-mapped with a bi-cubic
polynomial to correct for the focal plane curvature and to map the images
onto the $I$-band plate scale.  The sky
was removed by subtracting a smoothed fit to the minima in the sky.
The images were then summed with cosmic-ray rejection, with images showing
high extinction after correction for airmass being rejected from the sum.  
The resulting images
had seeing of $0\farcs 7$ in $R$-band and $0\farcs 8$ in $I$-band for RXJ
1716$+$67 and $0\farcs 7$ in $R$-band and $0\farcs 6$ in $I$-band for MS
1137$+$66.  These images, along with some shallower $B$-band images 
obtained from the UH 88'' telescope, have been combined into the three 
color images shown in Figure 1 (plate xx: MS 1137$+$66) and Figure 2 
(plate xx: RXJ 1716$+$67).

Photometric calibration was performed using the standard stars of Landolt
\markcite{r10} (1992).
The 1$\sigma$ surface brightness limits of the summed $R$-band and $I$-band
images for a seeing-sized disc are 29.4 and 26.6 for MS 1137$+$66 and 29.0 
and 26.8 for RXJ 1716$+$67.

The first step in the weak lensing analysis was detecting and measuring the
positions and luminosities of all the objects in the field.
We used a hierarchical peak finding algorithm to detect objects (Kaiser, 
Squires, and Broadhurst \markcite{r9} 1995), which provided
both centroid positions and a rough measure of the object size.
The luminosity of each object was measured using a circular aperture with
radius equal to three times the smoothing radius at which the object was
first detected. The catalog of objects was then used to mask the summed image.
Small, residual bias around each object left by the sky subtraction was then
measured in the masked image and subtracted.  Noise peaks and non-splitable
groups of objects were removed by rejecting extremely small and large
objects, as well as objects with abnormally high eccentricities.  With
very few exceptions, all objects detected in the I-band images were also
detected in the much deeper R-band images.  The I-band magnitudes were
calculated using an aperture $r_{ap} = 3r_g$, where $r_g$ is the smoothing
scale at which the object is first detected in the R-band image.  The
resulting I-band magnitudes agree with the magnitudes from the initial
detection in the I-band images to within 0.1 mag rms.  For purposes of color
selection, it is assumed that all objects present in the R-band images but
not detected in the I-band images are bluer than objects of the same R-band
magnitude with a detection in the I-band images.

The ellipticities of each object were measured using optical polarizations
$e_\alpha = \{I_{11} - I_{22}, 2I_{12}\}/(I_{11} + I_{22})$ formed from the
quadrupole moments $I_{ij} =
 \int d^2\Theta W(\Theta)\Theta_i\Theta_jf(\Theta)$
where $f$ is the flux density and $W(\Theta)$ is a Gaussian weighting function
of a scale equal to the smoothing radius at which the object was first
detected.  A low order polynomial
fit to a sample of bright but unsaturated stars ($\sim 50$ distributed across 
the field) was used to correct the ellipticities of the objects for psf
variations across the field (spherical aberrations and coma in the focal
plane, wind shake on the telescope, etc.).  The resulting ellipticities should
average to the object ellipticity, possibly sheared by the gravitational
lens, smeared by a circular psf.  Stars were removed from the catalog by using
half-light radii of the objects.

Given that the smeared ellipticities reduce the strength of any lensing
signal present, the next step is to boost the ellipticities of the galaxies
to remove the effects of the psf smearing.  This boost factor was
calculated using the shear polarizability $P_\gamma = P_{sh} - P_{sm}
*P^\star_{sh}/P^\star_{sm}$ (LK), where $P_{sh}$ and $P_{sm}$ are as 
defined in Kaiser, Squires, and Broadhurst (1995), and $\star$ denotes 
the values for stellar objects.  The
individual values of $P_\gamma$ are rather noisy for small galaxies, but
the mean value, $<P_\gamma>$, varies smoothly as a function of galaxy size.
Thus by binning the galaxies by size an estimate of the shear 
$\hat{\gamma}_\alpha = e_\alpha/<P_\gamma>$ was calculated.  This shear
is then related to the dimensionless surface density $\kappa$ by
$\gamma _\alpha  = \onehalf\{\phi_{,11} - \phi_{,22}, 2\phi_{,12}\}$ where 
$\kappa = \Sigma/\Sigma_{crit} = \onehalf\nabla^2\phi$ and where the
critical density $\Sigma_{crit} = 4\pi Gc^{-2}D_lD_{ls}D_s^{-1}$.
Because of the $D_{ls}D_s^{-1}$ factor in $\Sigma_{crit}$, the mass 
obtained  from $\kappa$ is dependent on the redshift of the source galaxies 
used in the 
weak lensing analysis.  In this paper we will be assuming that the background
galaxies lie on a sheet at $z=2.0$.  If, instead, the background galaxies were
to lie on sheets at $z=1.0$, $z=1.5$, or $z=3.0$, the measured masses would be
$3.0 \times$, $1.3 \times$, or $0.8 \times$ the quoted mass.
It should be noted that the intrinsic ellipticities of the galaxies cannot be
removed from $\hat{\gamma}_\alpha$ measurements, and are the dominant
source of noise in the analysis.

We used two different methods to derive $\kappa$ from the measured shears of
the background galaxies.  The first was the unbiased inversion algorithm
of Squires and Kaiser \markcite{r14} (1996).  This technique, while producing 
two dimensional
maps of the surface density, is only able to determine $\kappa$ to within an
unknown constant.  The second technique is that of aperture mass densitometry,
which measures the mass interior to a given radius (Fahlmen \etal
\markcite{r5} 1994).
The traditional statistic of aperture mass densitometry
\begin{eqnarray*}
\zeta(r_1) = \bar{\kappa}(r\leq r_1) - \bar{\kappa}(r_1<r\leq r_{max}) = \\
2 (1-r_1^2/r_{max}^2)^{-1} \int_{r_1}^{r_{max}} <\gamma_T> d\ln r
\end{eqnarray*}
provides a lower bound on $\bar{\kappa}$ interior to radius $r_1$.  However,
because this statistic subtracts $\bar{\kappa}$ of the annulus outside $r_1$,
the final measured mass $M(<r_1) = \pi r_1^2\zeta(r_1)\Sigma_{crit}$ depends
not only on the strength of the detected lensing but also on the mass
profile of the cluster.  Because of the large numbers of background galaxies
in our deep images, we have been able to use a slightly altered statistic
\begin{eqnarray*}
\zeta_c(r_1) = \bar{\kappa}(r\leq r_1) - \bar{\kappa}(r_2<r\leq r_{max}) =
2 \int_{r_1}^{r_2} <\gamma_T> d\ln r\\ + 2 (1-r_2^2/r_{max}^2)^{-1}
\int_{r_2}^{r_{max}} <\gamma_T> d\ln r
\end{eqnarray*}
which allows the subtraction of a constant $\bar{\kappa}(r_2<r\leq r_{max})$
for a fixed inner annulus radius $r_2$ for all apertures in the measurement,
thereby eliminating the dependence of the statistic on the mass 
profile of the cluster.  The measured mass profile $M = \pi r1^2
\zeta_c(r_1) \Sigma_{crit}$ is now the mass of the cluster minus an unknown,
but presumably small, constant times $r^2$.  In
both statistics the tangential shear is $<\gamma_T> = \int \gamma_T d\phi / 
2\pi $, where $\gamma_T = \gamma_1 \cos 2\phi + \gamma_2 \sin 2\phi$ and
$\phi$ is the azimuthal angle with respect to some chosen center (which we
take to be the peak of the smoothed light distribution for each cluster).

Both clusters' catalogs of objects were then divided into several subcatalogs 
by magnitude and color of the galaxies.  As was the case with MS 1054$-$03 in 
LK, both MS 1137$+$66 and RXJ 1716$+$67 had the strongest weak lensing 
signal from the catalogs containing the faintest, bluest galaxies, 
specifically  $23 < R < 27$ and $R-I < 1.3$.  The discussion of the results 
below come from the subcatalogs of the faint blue galaxies.

\section{Cluster Properties}
\subsection{MS 1137$+$66}
This cluster is easily identifiable as the compact group of red galaxies 
in the center of the field, seen in Figure 1 (plate xx).
Using a 500 $h^{-1}$ kpc aperture and a galaxy color selection of 
$1.3 < R-I < 2.0$ to isolate the cluster galaxies, an over-density of 
$56\pm6$ galaxies above the background level was determined from the 
galaxy density in the field outside the aperture. This corresponds to 
an Abell class 3 cluster richness (Bahcall \markcite{r2} 1981).
A smoothed image of the luminosities of the selected galaxies is shown in
Figure 3b.  The centroid of the cluster light peak is consistent with the
position of the BCG.  The total magnitude for all selected galaxies inside
a 500 $h^{-1}$ kpc aperture centered on the BCG, minus
a background determined from the galaxies outside the aperture, is $R = 
18.1\pm0.2$, which corresponds to a cluster luminosity 
$L_V = 9.1\pm 1.7 \times 10^{11} h^{-2} L_{V\sun}$. 

As can be seen in Figure 3a, MS 1137$+$66 has a mass distribution as compact
as the light (Figure 3b) and galaxy distribution.  The small peak just
north of the cluster is centered around three galaxies with colors similar
to the cluster galaxies, but of unknown redshift.  The minimum mass enclosed
in a 500 $h^{-1}$ kpc radius centered on the BCG (coincident with the 
center of the mass peak in the 2-d reconstruction), calculated using aperture
densitometry, is $2.45\pm0.8 \times  10^{14} h^{-1} M_\odot$.
The small northern peak is included in this estimate, but contributes less
than $10\%$ of the total mass detected.  Combining this mass estimate with the
cluster luminosity estimate from \S 3 gives $M/L_V = 270\pm100 h$, which
is similar to that found for MS 1054$-$03 (LK).  

The ultra-deep Keck image has also revealed a system of giant arcs present
in the core of MS 1137$+$66.  The brightest arc is located 5\farcs 5 
(22 $h^{-1}$ kpc) south-west of the BCG and has a surface brightness of $R = 
26.5$ mag$/$arcsec$^2$.  A second arc is located 
18\farcs 0 (73.6 $h^{-1}$ kpc)
north-east of the BCG and has a surface brightness of $R = 27.3$ 
mag$/$arcsec$^2$.  There is a potential counter arc with the same surface
brightness located 18\farcs 5 (75.7 kpc $h^{-1}$) south of the BCG.  There are
also two more arc candidates, all of which are marked in Figure 1 (plate xx).
No redshifts have yet been obtained for any of these arcs.  It is interesting
to note, however, that the outer giant arc is located presumably on the circle
at which $\bar{\kappa} = 1$ for the redshift of the arc.  The weak lensing
result, when extrapolated into this radius, has a $\bar{\kappa} = 
0.73\pm 0.09$.  This implies that the source galaxy redshift of the outer
giant arc is higher than the typical redshifts of the faint background
galaxy (FBG) population over the magnitudes which we have measured. 
The outer giant arc can be used to measure the mass of MS 1137$+$66 interior 
to its radius unambiguously, and, because of the large radius of the arc
from the cluster core, serve as an anchor point to the masses 
measured by the weak lensing.

\subsection{RXJ 1716$+$67}
Unlike MS 1137$+$66, RXJ 1716$+$67 does not appear to be a well formed 
cluster. As can be seen in Figure 2 (plate xx), the BCG is located on the 
north-western edge of the main group of galaxies, and there is a smaller
clump of galaxies to the north-east of the main group.  A filament of galaxies
extends north-east of the cluster, disappearing off the edge of the image.
A smoothed image of the luminosities of galaxies with $1.3 < R-I < 2.0$,
shown in Figure 3d, has the main peak in the cluster luminosity located
8\arcsec  southeast of the BCG.  In a 500 $h^{-1}$ kpc aperture centered on 
the peak of the cluster luminosity, RXJ 1716$+$67 has an over-density of 
galaxies of $52\pm7$, corresponding to Abell class 3 (Bahcall \markcite{r2} 
1981), and a cluster magnitude of $R = 17.9\pm0.2$ which 
corresponds to $L_V = 1.38\pm 0.26 \times 10^{12} h^{-2} L_{V\sun}$.  

While much of the fine structure seen in the smooth cluster light image
(Figure 3d) for RXJ 1716$+$67 is lost in the noise of the mass reconstruction
(Figure 3c), the mass of the northeastern group of galaxies is clearly 
separated from the mass of the main cluster.  Further, while the mass peak 
of the northeastern group is consistent with the center of light and the 
galaxy distribution of the group, the center of mass of the main cluster 
is located 27\arcsec (111 $h^{-1}$ kpc) east and 11\arcsec (45 $h^{-1}$ kpc) 
north of the BCG.  This offset is in the same direction but far larger than 
the offset seen in the center of the light peak of the cluster.  The minimum
mass enclosed in a 500
$h^{-1}$ kpc radius aperture centered on the center of mass for the main 
cluster, calculated using aperture densitometry, is $2.6\pm 0.9 \times  
10^{14} h^{-1} M_\odot$.  This includes both the mass of the main cluster 
and the smaller northeastern group of galaxies.  Combining this with the 
cluster luminosity estimate of \S 3 gives $M/L_V = 190\pm 70 h$.

\section{Discussion}
The mass profiles of both clusters are plotted in Figure 4, along with 
``universal'' CDM profiles of Navarro, Frenk, and White \markcite{r12} 
(1996). While the CDM profiles are not robust
fits (both parameters in the model can be adjusted to a small extent without
affecting the goodness of fit), they can be used as a general
comparison with mass profiles of lower redshift clusters.  In both clusters,
the observed masses are similar to those measured in lower redshift clusters
(Squires \etal \markcite{r15} 1996, Fischer \etal \markcite{r6} 1997), but
MS 1137$+$66 is much more compact than usual (Navarro \etal 1996).  

We believe that this compactness may be an artifact
of projecting 3D filamentary structures onto a 2D plane and that the 
compact profile of MS 1137$+$66 can be explained
if we are looking down the length of a filamentary structure.
This would cause all the mass of the central filament to be
projected as a large, steeply sloped 2D surface potential, 
and would not seriously affect 
the strength of the weak lensing provided the tube has a length smaller than
the typical background galaxy -- 2D cluster center distance.  We do not, 
however,
have enough redshifts on cluster galaxies to determine if this is indeed
the case.

Finally, these results show that a substantial part of the FBG population must
lie at $z > 1$.  Because of the dependence of $\bar{\Sigma}_{crit}$ on both 
the redshift of the lensing cluster and the redshifts of the background 
galaxies, we believe that the redshift distribution of the FBG population 
can be measured by using a number of clusters at varying redshifts.  The
outer giant arc in MS 1137$+$66 will prove to be a valuable tool in this
measurement.  The radius of the arc implies that the arc is being strongly
lensed primarily by the core of the cluster, and any nearby galaxies will
provide only a minor perturbation to the shape of the arc.
Thus, once the redshift of the outer arc is known, it will be possible to
obtain a mass measurement of the cluster interior to the arc with an uncertainty
much lower than both weak lensing and X-ray observations can obtain.  

\acknowledgments
We wish to thank Gillian Wilson, Mark Metzger, Lev Koffman, Len Cowie,
Dave Sanders, John Learned, Phil Fischer, James Bauer, and Neil Trentham for 
their help and advice.  We also wish to thank Harald Ebeling, Chris Mullis,
and Megan Donahue for sharing their X-ray data with us before publication.
This work was supported by NSF Grants AST-9529274 and AST-9500515, NASA-STScI
grant GO-540201-93A, and ASI grants ARS-94-10 and ARS-96-13.

\newpage
\figcaption{(Plate): $5\farcm 9\times 5\farcm 9$, 3 color image of 
MS 1137$+$66.  R, G, and B colors are
a 21000s $I$-band exposure from the UH88'' telescope, 8700s $R$-band 
exposure from the Keck II telescope, and a 12600s $B$-band exposure from 
the UH88'' telescope respectively.  All three colors are scaled with 
$\log ^{1/2}$ stretch. Inset in the lower right-hand corner is a greyscale
$\log (\log )$ close up of the cluster center in $R$-band.  Clearly evident 
are the two giant arcs (A and B).  Also present are the counter arc 
candidate (C), and two giant arc candidates (D and E).}

\figcaption{(Plate): $5\farcm 9\times 5\farcm 9$, 3 color image of
RXJ 1716$+$67.  R, G, and B colors are a 26100s $I$-band exposure from the
UH88'' telescope, a 7500s $R$-band exposure from the Keck II telescope, and
a 10800s $B$-band exposure from the UH88'' telescope respectively.  All three
colors are scaled with a $\log ^{1/2}$ stretch.  The BCG is located in the
lower center of the image, with a long filament leading out of the cluster
to the north-east.}

\begin{figure}
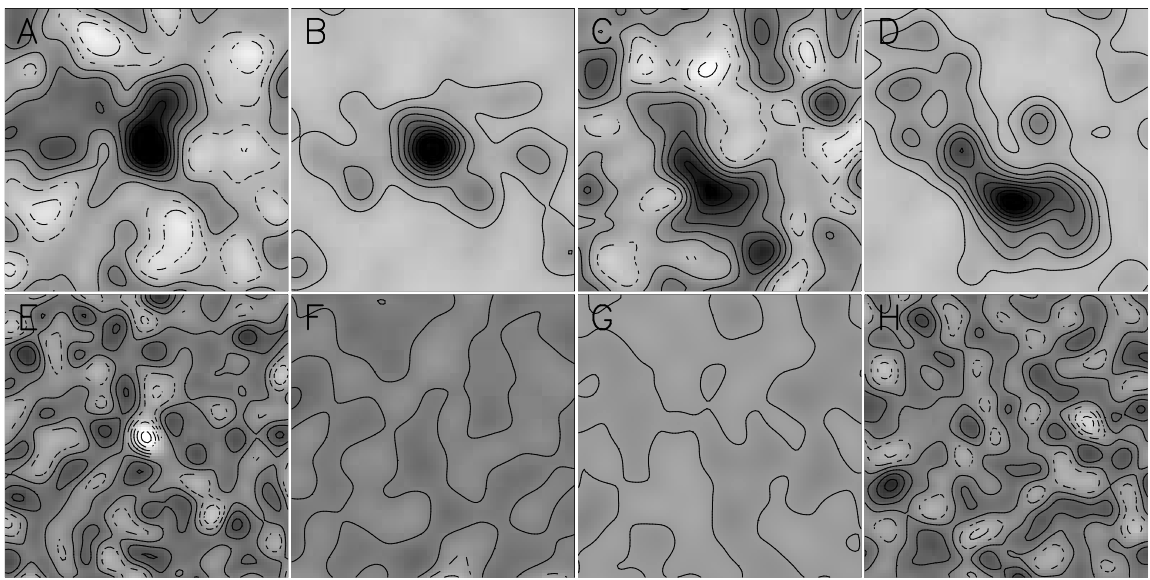

\vspace{2.5in}
\includegraphics{fig3a.ps}
\includegraphics{fig3b.ps}
\includegraphics{fig3c.ps}
\includegraphics{fig3d.ps}
\includegraphics{fig3e.ps}
\includegraphics{fig3f.ps}
\includegraphics{fig3g.ps}
\includegraphics{fig3h.ps}
\caption{The top four panels show the mass and light distribution for
MS 1137$+$66 (a and b) and RXJ 1716$+$67 (c and d).  Panel e is the Laplacian
$\nabla ^2\kappa$ of MS 1137$+$66, and panel f is $\nabla \times \nabla
\kappa $ of MS1137$+$66 (which should be 0 plus noise for a weak lensing
signal).  Panel g is the mass distribution from the background galaxies
in MS 1137$+$66, but with a randomized ellipticity orientation.  Panel
h is the $\nabla \times \nabla \kappa $ for the distribution in panel g.
In all cases, all panels which show a similar quantity have the same
absolute scale.}
\end{figure}

\begin{figure}
\vspace{6.0in}
\includegraphics{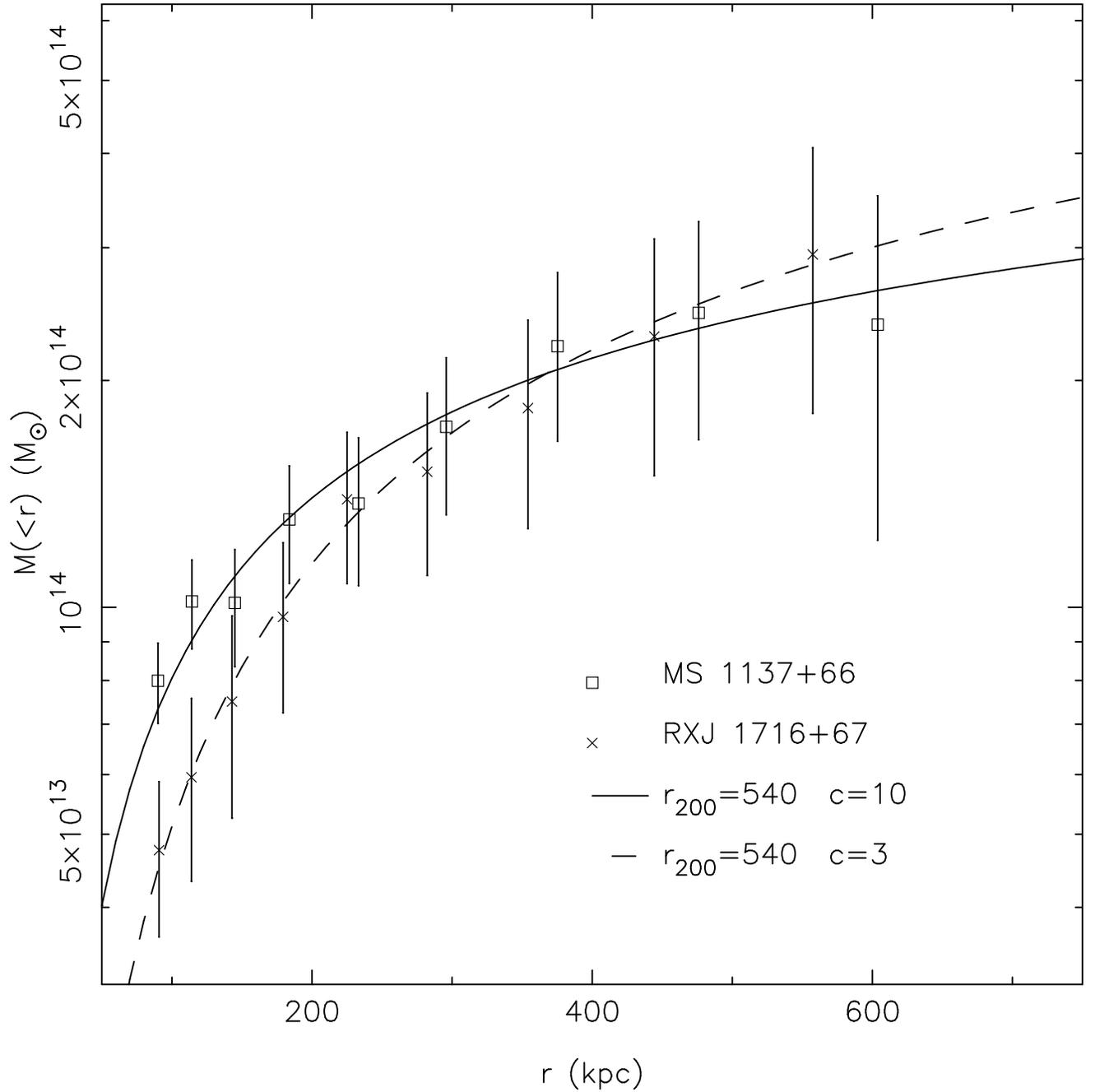}
\caption{Radial mass distribution of MS 1137$+$66 and RXJ 1716$+$67 from
aperture densitometry.  Also shown are some ``universal'' CDM profiles from
Navarro, Frenk, and White (1996).}
\end{figure}


\begin{references}
\reference{r1} Bahcall, N. A., Fan, X., \& Cen, R. 1997, \apjl, 485, L53.
\reference{r2} Bahcall, N. A. 1981, \apj, 246, 122.
\reference{r3} Crone, M. M., Evrard, A. E., \& Richstone, D. O. 1996, \apj, 
467, 489.
\reference{r4} Eke, V. R., Cole, S., \& Frenk, C. S. 1996, \mnras, 282, 263.
\reference{r5} Fahlman, G., Kaiser, N., Squires, G., \& Woods, D. 1994, 
\apj, 437, 56.
\reference{r6} Fischer, P., Bernstein, G., Rhee, G., \& Tyson, J. A. 1997, 
\aj, 113, 521.
\reference{r18} Gioia, I. M. \& Luppino, G. A. 1994, \apjs, 94, 583.
\reference{r17} Henry, J. P. 1997, \apjl, 489, L1.
\reference{r8} Henry, J. P., Gioia, I. M., Mullis, C. R., Clowe, D. I., 
Luppino, G. A., Boehringer, H., Briel, U. G., Voges, W., \& Huchra, J. P. 
1997, \aj, 114, 1293.
\reference{r9} Kaiser, N., Squires, G., \& Broadhurst, T. 1995, \apj, 449, 
460.
\reference{r10} Landolt, A. U. 1992, \aj, 104, 340.
\reference{r7} Luppino, G. A. \& Gioia, I. M. 1995, \apjl, 445, 77.
\reference{r11} Luppino, G. A. \& Kaiser, N. 1997, \apj, 475, 20 (LK).
\reference{r19} Morgan, J. A. 1995, PASPCS, 77, 129.
\reference{r12} Navarro, J. F., Frenk, C. S., \& White, S. D. M. 1996, 
\apj, 462, 563.
\reference{r13} Oke, J. B., Cohen, J. G., Carr, M., Cromer, J., Dingizian, 
A., Harris, F. H., Labrecque, S., Lucinio, R., Schaal. W., Epps, H., 
\& Miller, J. 1995, \pasp, 107, 375
\reference{r14} Squires, G. \& Kaiser, N. 1996, \apj, 473, 65.
\reference{r15} Squires, G., Kaiser, N., Babul, A., Fahlman, G., Woods, D., 
Neumann, D. M., \& Boehringer, H. 1996, \apj, 461, 572.
\reference{r16} Wilson, G., Cole, S., \& Frenk, C. S. 1996, \mnras, 282, 501.

\end{references}
\end{document}